# A lifestyle-based model of household neighbourhood location and individual travel mode choice behaviours

**Ali Ardeshiri (corresponding author)**

Institute for Choice, University of South Australia

Level 13, 140 Arthur Street

North Sydney, NSW 2060, Australia

Email: Ali.Ardeshiri@unisa.edu.au

**Akshay Vij**

Institute for Choice, University of South Australia

Level 13, 140 Arthur Street

North Sydney, NSW 2060, Australia

Email: vij.akshay@gmail.com

## Abstract

Issues such as urban sprawl, congestion, oil dependence, climate change and public health, are prompting urban and transportation planners to turn to land use and urban design to rein in automobile use. One of the implicit beliefs in this effort is that the right land use policies will, in fact, help to reduce automobile use and increase the use of alternative modes of transportation. Thus, planners and transport engineers are increasingly viewing land use policies and lifestyle patterns as a way to manage transportation demand. While a substantial body of work has looked at the relationship between the built environment and travel behaviour, as well as the influence of lifestyles and lifestyle-related decisions on using different travel modes and activity behaviours, limited work has been done in capturing these effects simultaneously and also in exploring the effect of intra-household interaction on individual attitudes and beliefs towards travel and activity behavior, and their subsequent influence on lifestyles and modality styles. Therefore, for this study we proposed a framework that captures the concurrent influence of lifestyles and modality styles on both household-level decisions, such as neighbourhood location, and individual-level decisions, such as travel mode choices using a hierarchical Latent Class Choice Model (LCCM).

The framework was empirically tested using travel diary data collected from households residing in the San Francisco Bay Area, United States. A six-class model was selected as the preferred specification to observe household residential neighbourhood choice behaviour. Coincidentally, based on both statistical measures of fit and behavioural interpretation, we also selected a six-class model as the preferred specification to observe the individual travel mode choice behaviour. Results from the models are presented which can have intriguing policy implications.

## 1. Introduction

Issues such as urban sprawl, congestion, oil dependence, climate change and public health, are prompting urban and transportation planners to turn to land use and urban design to rein in automobile use. As a result, increased interest in jobs-housing balance (Giuliano 1991, Cervero and Duncan 2006), smart growth (Handy 2005, Cervero and Duncan 2006), compact cities (Stead, Williams et al. 2000), transit oriented developments (Cervero 1994, Bernick and Cervero 1997), and new urbanism (Lund 2003, Ardeshiri and Ardeshiri 2011) have spawned concepts for researchers who diligently probe and dissect the many ways in which urban form, neighbourhood design, and the overall physical make-up of cities and regions shape how people travel. The relationship between urban form and sustainability is currently one of the most hotly debated issues on the international planning agenda. The way that cities should be developed in the future, and the effect that their form can have on resource depletion and social and economic sustainability, are central to this debate (Ardeshiri and Ardeshiri 2011).

One of the implicit beliefs in this effort is that the right land use policies will, in fact, help to reduce automobile use and increase the use of alternative modes of transportation (Handy 1996). While the empirical evidence so far provides support for this claim, closer examination of the evidence suggests that the relationships are more complex and the answers are not as clear as they may, at first, seem. Many studies have concluded that land use and the built environment measurements such as density, land-use diversity, design, accessibility and distance to transit, etc., influence the travel behaviour in statistically significant ways. Others have found no connection or a diverse relation as opposed by others (Reference). Zhou and Kockelman (2008) found that the built environment accounted for 58% to 90% of the total influence of residential location whereas in Bhat and Eluru (2009) this figure was 87% and in Raleigh, NC, Cao, Xu, and Fan (2010), 48% to 98% of the difference in vehicle miles driven was due to direct environmental influences, the balance being due to self-selection. Cao (2010) reported that, on average, neighbourhood type accounted for 61% of the observed effect of the built environment on utilitarian walking frequency and 86% of the total effect on recreational walking frequency. On the contrary for example Levinson and Kumar (1993) indicated that density and urban design did not explain much of the variation observed in transit usage or in Vehicle Mile Travel (VMT). This indicates that there is likely great heterogeneity within any population. Some people might be responsive, others may be very set in their ways regardless of changes in the built environment and thus the role of lifestyle and modality style come to play.

Researchers have examined the relationship between lifestyle and residential location choice (Ben-Akiva and Lerman 1985, Bagley and Mokhtarian 1999, Aeroe 2001, Waddell 2001, Walker and Li 2007), lifestyles and travel behaviour (Salomon and Ben-Akiva 1983, Simma and Axhausen 2001, Diana and Mokhtarian 2009, Kitamura 2009, Ohnmacht, Götz et al. 2009, Van Acker, Van Wee et al. 2010, Kuhnimhof, Buehler et al. 2012, Vij, Carrel et al. 2013, Xiong, Chen et al. 2015, Prato, Halldórsdóttir et al. 2016) and travel behaviour and land use (Bhat and Guo 2004, Pagliara and Wilson 2010, Sener, Pendyala et al. 2011, Zolfaghari, Sivakumar et al. 2012, Bhat 2015) in an aggregate and disaggregate level. However, the majority of the studies that examined the influence of lifestyles on different dimensions of transportation and land use behaviour have typically been conducted in isolation.

In this study we propose an integrated approach that recognizes the simultaneous influence of these higher order constructs on all dimensions of behaviour. In addition, the shift towards disaggregate models of decision-making has been seen as a significant step forward, contributing to the development of a comprehensive framework that recognizes the influence of land use patterns on the demand for transportation systems through the interplay between different dimensions of travel and activity behaviour. Thus, this study formulates a comprehensive model at a disaggregate level to investigate individual and household travel and activity behaviour that integrates travel demand and land use analysis through the construct of modality styles. This modelling framework benefits from offering a deeper understanding of decision-making procedure and greater predictive power than current models in practice. Furthermore, it will constitute an important component for urban and transport planners, city authorities and policy-makers to forecast how individuals arrive at decisions and how the city's form and shape might change in future.

The remainder of the paper is constituted as follows: section 2 provides an overview of the literature and background studies related to Land use and travel behaviour, lifestyle and modality style and models of group decision-making. Section 3 provides the methodological framework and a discussion on latent class choice models. Section 4 describe the data source for this study. Section 5 presents the empirical results of the latent class choice model. Section 6 presents conclusions and further directions.

## 2. Literature review and background studies

Section 2.1 reviews the substantial body of work on the relationship between the built environment and travel behaviour; Section 2.2 summarizes relevant findings from studies that have tested the influence of lifestyles and lifestyle-related decisions on different dimensions of individual and household travel and activity behaviours; and Section 2.3 concludes with a description of past work on discrete choice models of group decision-making, with a particular emphasis on studies in transportation and urban economics.

## 2.1 Land use and travel behaviour

Planners are increasingly viewing land use policies as a way to manage transportation demand. In one of the early studies Pushkarev & Zupan (1977), developed a set of land-use thresholds to financially justify the different types of transit investments, based on intermodal comparisons of transit unit costs and intercity comparisons of transit trip generation rates. Later in 1980 they developed six demand-based threshold criteria to determine the financial feasibility of fixed guide-way transit.

A potential way to moderate the transportation demand is by changing the built environment, which has become the most heavily researched subject in urban planning. In travel research, such influences have often been named with words beginning with D. The original "three Ds," coined by Cervero and Kockelman (1997), are density, diversity, and design, followed later by destination accessibility and distance to transit (Ewing and Cervero 2001, Ewing, Greenwald et al. 2009). Demand management, including parking supply and cost, is a sixth D, included in a few studies. While not part of the environment, demographics are the seventh D, controlled as confounding influences in travel studies. Below we only look at the literature on the original three D's.

*Density* is always measured as the variable of interest per unit of area. The area can be gross or net, and the variable of interest can be population, dwelling units, employment, building floor area, or something else. Population and employment are sometimes summed to compute an overall *activity density* per areal unit. Almost all studies have concluded that transit usage is influenced very significantly by residential density and there is a general tendency for less driving in higher density regions (Smith 1984, Dunphy and Fisher 1996, Schimek 1996a, Schimek 1996b, Cervero and Kockelman 1997, Salon 2006). On the contrary Ewing & Cervero (2010) concluded that with having other variables controlled, population and job densities are weakly associated with travel behaviour.

*Diversity* measures pertain to the number of different land uses in a given area and the degree to which they are represented in land area, floor area, or employment. Entropy measures of diversity, wherein low values indicate single-use environments and higher values more varied land uses, are widely used in travel studies. Several studies have looked at accessibility and land-use balance and led to the conclusions that land-use diversity generally reduce trip-rates and encourage non-auto travel in statistically significant ways (Cervero 1996, Cervero and Kockelman 1997, Kitamura, Mokhtarian et al. 1997, Kockelman 1997, Ewing and Cervero 2010, Ardeshiri 2014). Kockelman (1997) concluded that Land-use balance and accessibility are more relevant to travel-behaviour prediction than household and traveller characteristics commonly used for this purpose.

*Design* includes street network characteristics within an area. Street networks vary from dense urban grids of highly interconnected, straight streets to sparse suburban networks of curving streets forming loops and lollipops. Measures include average block size, proportion of four way intersections, and number of intersections per square mile. Design is also occasionally measured as sidewalk coverage (share of block faces with sidewalks); average building setbacks; average street widths; or numbers of pedestrian crossings, street trees, or other physical variables that differentiate pedestrian-oriented environments from auto-oriented ones. It has been argued that residents of neighbourhoods with grid-iron street designs and restricted commercial parking were found to average significantly less vehicle miles of travel (Kitamura, Mokhtarian et al. 1997, Kockelman 1997, Ewing and Cervero 2010). On the contrary Levinson and Kumar (1993) indicated that density and urban design did not explain much of the variation observed in transit usage or in Vehicle Mile Travel (VMT). Friedman et al. (1994) concluded that the daily trip-rate and auto driver trip-rate for households in the standard post-war suburbs with better designs to be significantly much higher than the corresponding rates for households in the traditional neighbourhood. Crane (1996) demonstrated that, under somewhat general conditions, changes in street network design will have an ambiguous effect on both the number of automobile trips and the total miles travelled by auto.

With regards to demographic characteristics, results for income has been seen consistence among other studies. For example, Dunphy and Fisher (1996) concluded that household travel by car increases with income, or in another study by Schimek (1996a) they found that relatively lower incomes and income growth in Toronto contributed to its higher observed use of public transit.

To conclude this section we argue that travel variables are generally inelastic with respect to change in measures of the built environment, but the combination effect of several such variables on travel could be substantial.

### 2.2 Lifestyles and modality styles

The empirical evidence so far provided suggests that the relationships between land use and travel behaviour are complex, outcomes are different and answers are not as clear as they may, at first, seem. Having said that, an ongoing and parallel stream of research, incorporating more subjective, rather than objective, determinants has been conducted that looks at variables such as habits, attitudes and lifestyle. It is argued that, for example, congestion is not a problem but merely a symptom of a true problem which is the "lifestyle" selection of the residents (Kitamura 2009). The concept of lifestyle is important to travel behaviour because the automobile, the dominant mode of urban travel today, is basic to the urban resident's lifestyle. Despite the fact, there has been no commonly accepted definition of the term ''lifestyle'' in the field of travel behaviour analysis and demand forecasting. The lifestyle concept extends the scope of travel behaviour analysis and may possibly lead to improved predictive performance of forecasting models.

The term ''lifestyle'' as used in the literature has two meanings: (a) activity and time use patterns and (b) values and behavioural orientation. These two are interrelated, but a critical difference exists: lifestyle as activity patterns may change as an individual adapts to a change in the environment, whereas lifestyle as orientation is one that the individual attempts to maintain by modifying behavioural patterns and adapting to the change. Change in lifestyle as orientation takes place in the long term through changes in values, attitudes, and preferences. A review of the present literature reveals that the measure of behaviour used in the travel characteristics of various populations includes trip rate, travel time, travel cost, mode choice and trip distance. These travel characteristics are measured with a set of factors that are believed to constrain and direct individual's activity choice and lifestyle (Kitamura 2009). These factors are certain personal characteristics (e.g., sex, stage in the life cycle, and health status) and roles that society assigns to persons (for detailed literature review please refer to Kitamura, 2009).

More recently, among publications that have related certain aspects of a person's mobility decisions (e.g., residential location choice, automobile type and activity participation) to lifestyle variables are Kitamura et al. (1997), Krizek and Waddell (2002), Lanzendorf (2002),

Choo and Mokhtarian (2004), Johansson et al. (2006) and Ory and Mokhtarian (2009), who found correlations between lifestyle-related explanatory variables and transportation choices. The argument has been made that long-term decisions, such as residential location or motor vehicle ownership, are not exogenous determinants of a person’s short-term mobility decisions, but rather a proxy for underlying ‘‘true’’ explanatory variables related to lifestyle (see, for example, (Boarnet and Crane 2001, Cervero and Duncan 2002, Srinivasan and Ferreira 2002, Schwanen and Mokhtarian 2005, Walker and Li 2007).

These findings substantiate the claim that modality styles form part of a person’s mobility style, and thus lifestyle, and that everyday mode choice decisions are strongly correlated with higher level orientations. Furthermore, lifestyle changes with socioeconomic, institutional, and technological changes. Increasing real income, decreasing working hours, and new consumer technology all contribute to the ever-evolving lifestyle of urban residents. The seemingly ever-expanding consumer demand leads to new products and services, industries and institutions, and urban forms. In addition, most studies have typically focused on one dimension of behaviour (mode choice, vehicle ownership or residential location) despite the fact that each of these behaviours are different expressions of the same lifestyle. , i.e. “where a man lives is not different from how many cars he needs.”

To gain an understanding of lifestyle and to develop the capability to predict its changes in the future, it is necessary to examine changes that take place in various elements of urban life and see how these changes are related to changes in lifestyle and travel behaviour. Therefore, we propose a more integrated approach that captures the simultaneous influence of modality styles on multiple dimensions of behaviour.

### 2.3 Models of group decision-making

Many dimensions of travel and activity behavior, such as residential location, are made collectively at the level of the household. Traditional travel demand models have tended to focus on the individual as the decision-making unit, and the influence of household interdependencies has only been indirectly captured through the use of household characteristics as explanatory variables (see, for example, Bowman, 1998). Though sociodemographic variables denoting household structure and individual characteristics can adequately represent the role different individuals' play within a household, they potentially overlook the direct effect of intra-household interactions and group dynamics on individual travel behavior (Bhat and Pendyala 2005).

Past studies on group decision-making have adopted one of two broad approaches. Early representations of the effect of household interactions on observable behavior have built on Becker's seminal work on time allocation theory (Becker 1965). According to Becker's original formulation, the household may be abstracted as a single unit of decision-making with a common preference function. While the simplicity of the approach is appealing, it overlooks differences across household members in terms of individual preferences and relative influence, and the accompanying process of bargaining and compromise between household members that results in a common preference function for the household as a whole. It has been argued that the use of models that disregard the decision-making mechanism underlying household behavior can lead to incorrect inferences regarding the impact of policies seeking to influence behavior, and conversely, a greater understanding of how households make decisions can strengthen the design of the same policies (see, for example, (Alderman, Chiappori et al. 1995, Lundberg, Pollak et al. 1997, Vermeulen 2002, Adamowicz, Hanemann et al. 2005, Munro 2009).

In an attempt to overcome some of these drawbacks, studies on travel demand analysis in the last decade have sought to model explicitly the dynamic interplay between each of the different members that make up a household. The focus of most of these studies has been on the generation and allocation of household activities between household members (see, for example, (Gliebe and Koppelman 2005, Meister, Frick et al. 2005, Srinivasan and Bhat 2005, Kato and Matsumoto 2009, Roorda, Carrasco et al. 2009, Wang and Li 2009, Zhang, Kuwano et al. 2009) and the division of shared resources, such as a car, among household members (see, for example, (Golob, Kim et al. 1996, Arentze and Timmermans 2004, Petersen and Vovsha 2006, Roorda, Carrasco et al. 2009). These studies have repeatedly confirmed the presence of significant interaction effects between household heads, and the persistence of strong gender, income, employment and life- cycle effects on the pattern of allocation of activities and resources among household members.

However, the focus of this study is not on the generation and allocation of different household activities or the division of shared resources. Rather, we wish to explore the effect of intra-household interaction on individual attitudes and beliefs towards travel and activity behavior, and their subsequent influence on lifestyles and modality styles. An individual's preferences and choices are strongly shaped by the opinions and behaviors of the people around her (Thorndike 1938, Rose and Hensher 2004, Zhang, Kuwano et al. 2009), particularly when a choice is made collectively by a group of individuals, as in the case of a household. Studies on group decision-making within both travel behavior and the marketing sciences have sought to decompose household preferences as some function of the individual preferences of the household members and their relative influence on the decision-making process. For example, early work by Rao and Steckel (1991) formulates the utility derived by the group as a linear combination of the utility derived by each of its constituent members, weighted by the relative influence exerted by each member. Arora and Allenby (1999) extend their framework to allow the weights to vary by product attribute, thereby capturing individual influence on specific aspects of the shared preference function of the group. Work by Aribarg et al. (2002), Rose and Hensher (2004) and Hensher et al. (2008) extend this framework further to account explicitly for the bargaining process through which individuals may revise their preferences and/or concede to another member's.

Our work builds upon past research on group decision-making within the context of individual and household travel and activity behavior through the introduction of the household modality style construct. As mentioned previously, many of the dimensions of travel and activity behavior studied by travel demand analysts involve choices made at the level of the household. The preferences of the household are the outcome of a process of negotiation between the individuals that comprise the household and their respective preferences (Corfman 1991, Lee and Beatty 2002). In turn, the preferences of the individuals themselves are shaped by the preferences of other household members, and are therefore some reflection of the preferences of the household as a whole (Davis 1973, Menasco and Curry 1989). A comprehensive travel demand model must recognize the dialogue between individual and household preferences, or modality styles, that underlies observable behavior. However, unlike some of the studies cited in previous paragraphs that have been very detailed in their representation of the dynamics underlying group decision-making, we won't be as explicit. That being said, we will be relying on findings from these studies to develop a simpler framework that captures the reciprocal influence of individual and household modality styles on each other and concurrently on different dimensions of observable travel and activity behavior.

**Figure 1:** The proposed hierarchical latent class choice model of household residential neighbourhood choice and individual travel mode choices

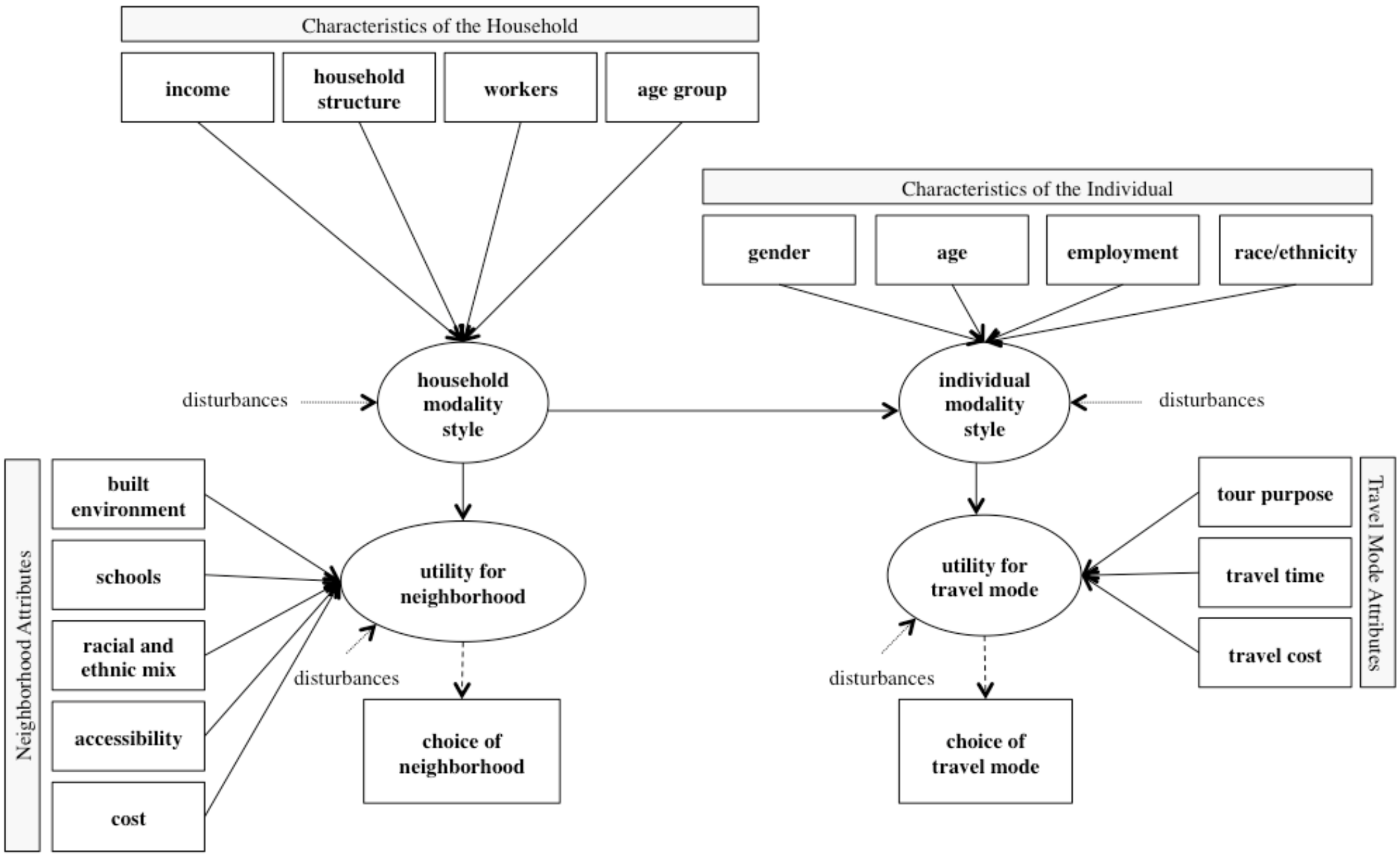

## 3. Methodological Framework

Existing travel demand modelling practice relies on sequentially nested logit models for a hierarchical representation of different choice dimensions. Lower dimensions, such as travel mode choice, are conditioned on purportedly higher dimensions, such as residential location, creating a vertical chain of inter-connected nests that in their totality represent an individual's travel and activity behaviour. While such a representation is convenient from the standpoint of estimation, it overlooks the concurrent influence of individual and household modality styles on all dimensions of an individual's travel and activity behaviour.

In developing a framework that captures the concurrent influence of lifestyles and modality styles on both household-level decisions, such as neighbourhood location, and individual-level decisions, such as travel mode choices, we propose using the hierarchical Latent Class Choice Model (LCCM) illustrated in Figure 1. Latent classes at the higher level represent household modality styles and latent classes at the lower level represent individual modality styles. We argue that households may be decomposed into discrete segments that differ in their predisposition towards different neighbourhood types and their sensitivity to different neighbourhood attributes. Similarly, household members may be decomposed into discrete segments themselves that differ in their awareness of and proclivity towards different travel modes, and their sensitivity, or lack thereof, to different level-of-service attributes of the transportation system. These differences are indicative of overarching differences in household and individual modality styles that concurrently influence neighbourhood choice and travel mode choices over time.

The proposed model framework comprises four components: (1) household class membership model; (2) neighbourhood choice model; (3) individual class membership model; and (4) travel mode choice models. We begin by describing the household class membership model, which predicts the probability that household $\mathrm{h}$ belongs to household modality style $\mathrm{r}$, and is formulated as a multinomial logit model:

$$\mathrm{P}(\mathrm{q}_{\mathrm{hr}} = 1) = \frac{\exp(\mathbf{z}_{\mathbf{h}}'\boldsymbol{\alpha}_{\mathbf{r}})}{\sum_{\mathbf{r}'=\mathbf{1}}^{\mathrm{R}} \exp(\mathbf{z}_{\mathbf{h}}'\boldsymbol{\alpha}_{\mathbf{s}'})} \qquad (1)$$

, where $\mathrm{q}_{\mathrm{hr}}$ equals one if household $\mathrm{h}$ belongs to modality style $\mathrm{r}$, and zero otherwise; $\mathbf{z}_{\mathbf{h}}$ is a vector of characteristics of household $\mathrm{h}$, such as income and household structure; $\boldsymbol{\alpha}_{\mathbf{r}}$ is a vector of parameters associated with the household's characteristics; and $\mathrm{R}$ denotes the number of household modality styles in the sample population. Note that $R$ must be determined exogenously, by estimating models with different numbers of classes and comparing estimation results in terms of both model fit and behavioural interpretation.

Residential location is subsequently conditioned on the household's modality style. Let $u_{hj|r}$ be the utility of neighbourhood j for household $h$, given that the household belongs to latent class $r$:

$$u_{hj|r} = \mathbf{x}_j'\boldsymbol{\beta}_r + \varepsilon_{hj|r} \tag{2}$$

, where $\mathbf{x}_j$ is a vector of attributes of neighbourhood j; $\boldsymbol{\beta}_r$ is a vector of class-specific parameters denoting sensitivities to each of these attributes; and $\varepsilon_{hj|r}$ is the stochastic component of the utility specification, assumed to be i.i.d. Extreme Value across households, neighbourhood s and classes with location zero and scale one. Assuming that all individuals are utility maximizers, the class-specific neighbourhood choice model may be formulated as follows:

$$P(y_{hj} = 1 | q_{hr} = 1) = \frac{\exp(\mathbf{x}_j'\boldsymbol{\beta}_r)}{\sum_{j' \in \mathbf{J}_r} \exp\left(\mathbf{x}_{j'}'\boldsymbol{\beta}_r\right)} \tag{3}$$

, where $y_{hj}$ equals one if household $h$ resides in neighbourhood j, and zero otherwise; and $\mathbf{J}_r$ denotes the set of all neighbourhood s considered by households belonging to class $r$. Heterogeneity in the decision-making process is captured by allowing both the taste parameters $\boldsymbol{\beta}_r$ and the consideration set $\mathbf{J}_r$ to vary across modality styles. For example, some classes might be more sensitive to land use variables, such as density and diversity, and others might base their decision on other variables, such as quality of schooling or crime rate. Equation (3) may be combined iteratively over all neighbourhood s in the set $\mathbf{J}_r$ to yield the following conditional probability of observing the vector of neighbourhood location choices $\mathbf{y_h}$ for household h:

$$f_y(\mathbf{y_h}|q_{hr}=1)=\prod_{j'\in \mathbf{J_r}}\left[P\left(y_{hj}=1|q_{hr}=1\right)\right]^{y_{hj}} \tag{4}$$

Moving on to the individual modality styles construct, the probability that individual $n$ from household $h$ has modality style $s$, conditional on the household belonging to modality style $r$, is assumed to be multinomial logit, and can be expressed as:

$$P(g_{hns}=1|q_{hr}=1)=\frac{\exp(\mathbf{w'_{hn}}\boldsymbol{\gamma}_{\mathbf{rs}})}{\sum_{\mathbf{s'=1}}^{S}\exp(\mathbf{w'_{hn}}\boldsymbol{\gamma}_{\mathbf{rs'}})} \tag{5}$$

, where $g_{hns}$ equals one if individual $n$ from household $h$ belongs to modality style $s$, and zero otherwise; $\mathbf{w_{hn}}$ is a vector of characteristics of the individual, such as age and gender; $\boldsymbol{\gamma}_{\mathbf{rs}}$ is a vector of model parameters associated with the individual's characteristics; and $S$ denotes the number of individual modality styles in the sample population. As was the case with the number of household modality styles $R$, note that $S$ too must be determined exogenously, by estimating models with different numbers of classes and comparing estimation results in terms of both model fit and behavioural interpretation.

The interaction between household members is captured by conditioning the model parameters $\boldsymbol{\gamma}_{\mathbf{rs}}$ on the modality style of the household as a whole. In other words, individuals with identical demographic characteristics could potentially belong to very different modality styles, depending on the household's modality style. For example, young men belonging to households that are collectively car-oriented may be more likely to be car-oriented themselves, and similarly, young men belonging to households that are more multimodal may be more likely to be multimodal themselves. Differences in the propensity of the same demographic subgroup to belong to different individual modality styles, due to differences in household modality styles, can offer insight on the influence of household interactions on individual lifestyles.

And finally, individual travel mode choices are conditioned on the individual's modality style. Let $v_{hndtk|s}$ be the utility of travel mode $k$ over tour $t$ for tour purpose $d$ and individual $n$ belonging to household $h$, given that the individual belongs to latent class $s$:

$$v_{hndtk|s} = \mathbf{c}'_{\mathbf{hndtk}}\boldsymbol{\lambda}_{\mathbf{ds}} + \eta_{hndtk|s} \tag{6}$$

, where $\mathbf{c}_{\mathbf{hndtk}}$ is a vector of attributes of travel mode $k$ over tour $t$ for tour purpose $d$ and individual $n$ belonging to household $h$; $\boldsymbol{\lambda}_{\mathbf{ds}}$ is a vector of class-specific parameters denoting sensitivities to each of these attributes for tour purpose $d$ and class $s$; and $\eta_{hndtk|s}$ is the stochastic component of the utility specification, assumed to be i.i.d. Extreme Value across households, individuals, tour purposes, tours, travel modes and classes with location zero and scale one. Assuming that all individuals are utility maximizers, the class-specific travel mode choice model may be formulated as follows:

$$P(m_{hndtk} = 1 | g_{hns} = 1) = \frac{\exp(\mathbf{c}'_{\mathbf{hndtk}}\boldsymbol{\lambda}_{\mathbf{s}})}{\sum_{k' \in \mathbf{K}_{\mathbf{ds}}} \exp\left(\mathbf{c}'_{\mathbf{hndtk'}}\boldsymbol{\lambda}_{\mathbf{s}}\right)} \tag{7}$$

, where $m_{hndtk}$ equals one if individual $n$ from household $h$ chooses travel mode $k$ over tour $t$ for tour purpose $d$, and zero otherwise; and $\mathbf{K}_{\mathbf{ds}}$ denotes the set of all travel modes considered by individuals belonging to class $s$ for tours with purpose $d$. We introduce the subscript $d$ to account for the fact that we have separate class-specific mode choice models for mandatory and non-mandatory tours. Heterogeneity in the decision-making process is captured by allowing both the taste parameters $\boldsymbol{\lambda}_{\mathbf{ds}}$ and the consideration set $\mathbf{K}_{\mathbf{ds}}$ to vary across modality styles. For example, individuals belonging to a particular class might always drive, regardless of the level-of-service of other travel modes, whereas individuals belonging to a different class might consider all travel modes, but have a high value of time that similarly predisposes them towards the car.

Equation (7) may be combined iteratively over travel modes, tours and tour purposes to yield the following conditional probability of observing the vector of choices $\mathbf{m_{hn}}$:

$$f_{\mathbf{m}}(\mathbf{m_{hn}}|g_{hns} = 1) = \prod_{d=1}^{D} \prod_{t=1}^{T_{hnd}} \prod_{k' \in \mathbf{K_{ds}}} [P(m_{hndtk} = 1|g_{hns} = 1)]^{m_{hndtk}} \quad (8)$$

, where $D$ denotes the number of tour purposes, two in our case; and $T_{hnd}$ denotes the number of observed tours for individual $n$ from household $h$ for tour purpose $d$. Equation (8) may be marginalized over the distribution of individual modality styles to yield the probability function of observing the vector of choices $\mathbf{m_{hn}}$, conditional on the household's modality style. This may subsequently be combined iteratively over all individuals belonging to the household, multiplied by equation (4), marginalized over the distribution of household modality styles, and iteratively combined over all households in the sample population to yield the following likelihood function:

$$L(\mathbf{y}, \mathbf{m}) = \prod_{h=1}^{H} \sum_{r=1}^{R} P(q_{hr} = 1) f_{\mathbf{y}}(\mathbf{y_h}|q_{hr} = 1) \prod_{n=1}^{N_h} \sum_{s=1}^{S} f_{\mathbf{m}}(\mathbf{m_{hn}}|g_{hns} = 1) P(g_{hns} = 1|q_{hr} = 1) \quad (9)$$

, where $H$ denotes the number of households in the sample population; and $N_h$ denotes the number of individuals that belong to household $h$. The unknown model parameters $\boldsymbol{\alpha}$, $\boldsymbol{\beta}$, $\boldsymbol{\gamma}$ and $\boldsymbol{\lambda}$ may be estimated by maximizing the likelihood function given by equation (9). However, simultaneous estimation of the full model proved to be computationally challenging. The model reported in this paper was ultimately estimated sequentially through a three-step procedure: (1) estimate an LCCM of household neighbourhood choice behaviour to determine $\mathrm{R}$, the number of household modality styles, $\boldsymbol{\beta}$, the vector of class-specific model choice model parameters, and $\boldsymbol{\alpha}$, the vector of household modality style class membership model parameters; (2) estimate an LCCM of individual travel mode choice behaviour to determine $\mathrm{S}$, the number of individual modality styles, and $\boldsymbol{\lambda}$, the vector of class-specific model choice model parameters; and (3) use the estimates for $\boldsymbol{\alpha}$ and $\boldsymbol{\beta}$ from the first step and $\boldsymbol{\lambda}$ from the second, and maximize equation (9) for $\boldsymbol{\gamma}$. All models and sub-models were estimated in Python using the Expectation-Maximization algorithm and an implementation of the BFGS algorithm contained in the SciPy library (Jones, Oliphant et al. 2001).

The modelling approach is exploratory in that both the number of household and individual modality styles and the behaviour of each modality style emerge naturally from the process of testing different model specifications. As we'll demonstrate in the subsequent case study, differences in taste parameters across the neighbourhood and travel mode choice class-specific models, and how these differences are correlated through the individual and household modality styles constructs, can offer insight on the extent to which households self-select into different built environments and the extent to which the built environment influences travel behaviour.

## 4. Data

The data used for investigating the proposed hierarchical latent class choice model of household residential neighbourhood choice and individual travel mode choices is drawn from the 2010-2012 California Household Travel Survey (CHTS). CHTS is a multi-modal study of the demographic and travel behaviour characteristics of residents across the entire state of California. The main objective of this survey is to be able to apply the data to develop and update transportation models in order to meet statutory requirements of both Federal (air quality analysis) and State Sustainable Communities and Climate Protection Act of 2008 (SB 375), Assembly Bill 32 (AB 32) and Senate Bill No. 391 (SB 391) to reduce greenhouse gas emissions through coordinated transportation and land use planning with the goal of more sustainable communities.

CHTS 2010-12 is the largest single regional household travel survey ever conducted in the United States. In total 42,500 households from 58 counties, plus portions of three adjacent counties in Nevada were sampled, using combination of computer assisted telephone interviewing (CATI), online, and three types of global positioning systems (GPS) devices--wearable, in-vehicle and in-vehicle plus an on-board diagnostic (OBD) unit. The survey sampling plan was designed to ensure an accurate representation of the entire population of the State. Thus, an address-based sampling frame approach was used to randomly sample all residential addresses that receive U.S. Mail delivery. Its main advantage is its reach into population groups that typically participate at lower-than-average levels, largely due to coverage bias (such as households with no phones or cell-phone only households). For efficiency of data collection, NuStats matched addresses to telephone numbers that had a listed name of the household appended to the sampled mailing addresses. This sampling frame ensured coverage of all types of households irrespective of their telephone ownership status, including households with no telephones (estimated at less than 3% of households in the U.S.).

In order to better target hard-to-reach groups, the address-based sample was supplemented with samples drawn from the listed residential frame that included listed telephone numbers from working blocks of numbers in the United States for which the name and address associated with the telephone number were known. The "targeted" Listed Residential sample, as available from the sampling vendor, included low-income listed sample, large-household listed sample, young population sample, and Spanish-surname sample (to name a few). As expected, this sample was used to further strengthen the coverage of hard-to-reach households. The advantage of drawing sample from this frame is its efficiency in conducting the survey effort—being able to directly reach the hard-to-reach households and secure their participation in the survey in a direct and active approach.

For the purposes of this study, we only looked at households residing in the San Francisco Bay area (henceforward the Bay Area), which includes nine counties: Alameda, Contra Costa, Marin, Napa, San Francisco, San Mateo, Santa Clara, Solano, and Sonoma. In all, we have data from 8,208 households. For each household, we know the census tract location of their current residence. Secondary data related to characteristics of each of the 1577 census tracts in the Bay Area was extracted from the 2013 TIGER/Line Shapefiles and the 2009-2013 American Community Survey (ACS) 5-year estimates data. The reader is referred to Appendix (A) for an exhaustive list of the variables used in our model specification.

Individuals from participating households were requested to provide additional information on all in-home and out-of-home activities over a one-day period. We processed this information into home-based tours and subsequently classified the tours as mandatory (those that include a work or school stop) or non-mandatory (all other). The resulting dataset includes 9,762 mandatory tours and 17,292 non-mandatory tours made by 17,680 individuals from the 8,208 households. For each tour, five feasible travel mode alternatives are defined: private vehicle, private transit, public transit, bike, and walk. Private vehicle refers to cases where the individual used a motorized vehicle owned by themselves (or someone they know) as a driver or a passenger. Private transit includes the use of travel modes such as taxis, Uber, car share, rental cars and private shuttles. Public transit includes modes such as buses, trains, ferries, etc. Level-of-service attributes, namely travel times and costs, for each of the six travel modes are constructed using network skims from the SF MTC for 2000 and 2010, generated using the same travel demand model. We were unable to decompose travel time into its constituent elements, such as in-vehicle time and waiting time, as this information was not available to us.

## 5. Estimation Results

As described in Section 3, we used a three-step procedure to determine the final model specification.

First, we estimated multiple models of individual travel mode choice behaviour, to determine both the number of individual modality styles and the class-specific mode choice models for each of these modality styles. Table 1 reports summary statistics for different model specifications. Coincidentally, based on both statistical measures of fit and behavioural interpretation, we selected the six-class model as the preferred specification in this case as well. Tables 3 and 4 report estimation results for the class-specific choice models for mandatory and non-mandatory tours, respectively.

Second, we estimated multiple models of household residential neighbourhood choice behaviour, to determine the number of household modality styles, and the class membership model and class-specific neighbourhood choice models for each of these modality styles. Table 2 reports summary statistics for each of the different model specifications corresponding to a particular number of classes, deemed the best for that number of classes. We selected the six-class model as the preferred specification. However, the reader will note that six classes were the most we explored. This was in part due to computational limitations. In subsequent work, we intend to evaluate model specifications with more classes. Tables 5 and 6 report estimation results for the class membership model for household modality styles and the class-specific choice models for residential neighbourhood, respectively.

And third, we use the full likelihood function to re-estimate the class-membership model corresponding to individual modality styles, conditional on the household modality style, constraining all other model parameters to be equal to estimates from the first two steps. Table 7 reports estimation results for the class membership model.

Based on statistical measures of fit, we select the six-class LCCM for both the individual and household modality styles as the preferred model specification.

**Table 1.** Summary statistics for different specifications for the sub-model of individual modality styles and travel mode choices

| Model | Parameters | Log-likelihood | $\bar{\rho}^2$ | AIC | BIC |
|---|---|---|---|---|---|
| Two classes | 37 | -9,784 | 0.718 | 19,641 | 19,941 |
| Three classes | 59 | -9,682 | 0.720 | 19,482 | 19,960 |
| Four classes | 75 | -9,619 | 0.721 | 19,388 | 19,996 |
| Five classes | 99 | -9,527 | 0.723 | 19,252 | 20,054 |
| Six classes | 127 | -9,477 | 0.724 | 19,207 | 20,236 |
| Seven classes | 155 | -9,460 | 0.724 | 19,229 | 20,485 |
| Eight classes | 173 | -9,446 | 0.723 | 19,239 | 20,641 |
| Nine classes | 197 | -9,442 | 0.723 | 19,279 | 20,875 |
| Ten classes | 218 | -9,441 | 0.722 | 19,318 | 21,084 |
| Eleven classes | 241 | -9,430 | 0.722 | 19,342 | 21,295 |

**Table 2.** Summary statistics for different specifications for the sub-model of household modality styles and residential neighbourhood choices

| Model | Parameters | Log-likelihood | $\bar{\rho}^2$ | AIC | BIC |
|---|---|---|---|---|---|
| Two classes | 42 | -36,248 | 0.043 | 72,580 | 72,882 |
| Three classes | 72 | -35,604 | 0.059 | 71,353 | 71,869 |
| Four classes | 102 | -35,252 | 0.068 | 70,709 | 71,441 |
| Five classes | 132 | -35,037 | 0.073 | 70,339 | 71,286 |
| Six classes | 162 | -34,794 | 0.078 | 69,911 | 71,074 |

## 5.1 Individual modality styles estimates

As mentioned previously, the final specification for the individual travel mode choice behaviour sub-model identified six classes in the sample population that differ from each other in terms of the travel modes that they consider, their relative sensitivity to travel times and travel costs, and their demographic characteristics. Over subsequent paragraphs, we summarize some of the key attributes of each of these six classes. To ease comprehension, we've ordered the classes in terms of decreasing car dependence.

**Individual Class 1:** This class constitutes 30.9% of the sample population. Individuals belonging to this class deterministically choose private vehicle for both mandatory and non-mandatory tours. In other words, they are completely dependent on their cars to fulfil their mobility needs. Married adults with children are most likely to be in this class.

**Individual Class 2:** This class constitutes 22.5% of the sample population. Individuals belonging to this class consider all travel modes except private transit for both mandatory and non-mandatory tours. In terms of actual use though, the class is heavily dependent on the car, with mode shares of 88% and 93% for mandatory and non-mandatory tours, respectively. Individuals belonging to the class are insensitive to travel costs, but have high mean elasticities to travel times. Similar to class 1, married adults with children are most likely to be in this class.

**Table 3:** Parameter estimates (and t-statistics) for the class-specific travel mode choice model for mandatory tours

| Model | Complete car dependence | Partial car dependence | | Multimodal | Low car dependence | |
|---|---|---|---|---|---|---|
| | Class 1 | Class 2 | Class 3 | Class 4 | Class 5 | Class 6 |
| *Alternative specific constants* | | | | | | |
| Private vehicle | 0.000 (-) | 0.000 (-) | 0.000 (-) | 0.000 (-) | 0.000 (-) | 0.000 (-) |
| Private transit | - | - | - | -2.802 (-13.987) | -1.667 (-2.478) | - |
| Public transit | - | 0.472 (1.496)* | 2.842 (5.464) | -1.439 (-5.375) | 4.979 (8.202) | 10.911 (19.663) |
| Bike | - | -1.422 (-9.345) | - | -5.4584 (-2.656) | 3.684 (9.924) | - |
| Walk | - | 1.245 (5.605) | 0.451 (1.352)* | -2.585 (-5.521) | 6.567 (12.630) | 11.326 (16.463) |
| *Level-of-service* | | | | | | |
| Travel time (minutes) | - | -0.0656 (-17.697) | -0.096 (-15.698) | -0.001 (-1.410)* | -0.087 (-13.648) | -0.061 (-13.66) |
| Travel cost ($) | - | - | -1.005 (-16.833) | - | -0.064 (-1.038)* | -0.532 (-6.720) |
| *Choice probability elasticities* | | | | | | |
| Travel time (minutes) | 0.000 | -11.236 | -18.439 | -0.171 | -10.553 | -8.460 |
| Travel cost ($) | 0.000 | 0.000 | -1.091 | 0.000 | -0.189 | -0.574 |
| *Marginal rates of substitution* | | | | | | |
| Value of time ($/hr) | - | ∞ | 5.731 | ∞ | 81.844 | 6.831 |

*Not Significant at 95% level

**T-test estimates are presented in the parenthesis.

**Individual Class 3:** This class constitutes 39.3% of the sample population. Individuals belonging to this class consider private vehicle, public transit and walk for both mandatory and non-mandatory tours. However, like Class 2, in terms of actual use, the class is almost equally dependent on the car, with mode shares of 94% and 79% for mandatory and non-mandatory tours, respectively. Unlike Class 2, this class is sensitive to both travel times and travel costs, and the value of time was estimated to be 5.7$/hr for mandatory tours and 19.3$/hr for non-mandatory tours. Individuals belonging to this class are younger (average age of 40 years, compared to 48 years for Class 2), have lower employment rates (46%, compared to 65% for Class 2), and higher school enrolment rates (34%, compared to 10% for Class 2).

**Individual Class 4:** This class constitutes 3.7% of the sample population. Individuals belonging to this class consider all travel modes for both mandatory and non-mandatory tours, though three in four mandatory tours and one in two non-mandatory tours are still made by private vehicle. Of the other modes, public transit is most popular for mandatory tours, with a mode share of 13%, and walk is most popular for non-mandatory tours, with a mode share of 34%. The class has low mean elasticities towards both travel times and costs, for both mandatory and non-mandatory tours, indicating that travel mode decisions are determined by factors other than these level-of-service attributes. The class has the highest average age (50 years) and the lowest rate of parenthood (11%), indicating that individuals belonging to this class likely lie later in their life cycles than other classes.

**Individual Class 5:** This class constitutes 1.6% of the sample population. Like Class 4, individuals belonging to this class consider all travel modes for both mandatory and non-mandatory tours. However, unlike Class 4, travel mode shares for private vehicle are significantly lower, with 38% of mandatory tours and 56% of non-mandatory tours being made by that mode. The class has high mean elasticities towards travel time and high values of time for both tour types. Bike is by far the most popular alternative mode of transport, accounting for 37% of mandatory tours and 23% of non-mandatory tours. This is the only class with a significant skew towards one gender: 78% of individuals belonging to this class are male.

**Table 4:** Parameter estimates (and t-statistics) for the class-specific travel mode choice model for non-mandatory tours

| Model | Complete car dependence | Partial car dependence | | Multimodal | Low car dependence | |
|---|---|---|---|---|---|---|
| | Class 1 | Class 2 | Class 3 | Class 4 | Class 5 | Class 6 |
| *Alternative specific constants* | | | | | | |
| Private vehicle | 0.000 | 0.000 | 0.000 | 0.000 | 0.000 | 0.000 |
| | (-) | (-) | (-) | (-) | (-) | (-) |
| Private transit | - | - | - | -3.318 | -4.658 | - |
| | | | | (-10.316) | (-5.022) | |
| Public transit | - | 0.472 | -1.621 | -1.322 | 1.348 | 2.784 |
| | | (0.982)* | (-2.660) | (-4.985) | (2.688) | (5.209) |
| Bike | - | -2.955 | - | -1.678 | 0.632 | -3.795 |
| | | (-13.602) | | (-8.996) | (3.631) | (-2.182) |
| Walk | - | 1.256 | 1.818 | 0.359 | 2.525 | 3.623 |
| | | (5.252) | (12.396) | (2.271) | (8.732) | (6.312) |
| *Level-of-service* | | | | | | |
| Travel time (minutes) | - | -0.141 | -0.047 | -0.008 | -0.069 | -0.059 |
| | | (-18.420) | (-22.389) | (-9.381) | (-14.237) | (-8.697) |
| Travel cost ($) | - | - | -0.146 | -0.014 | - | -0.478 |
| | | | (-1.452*) | (-0.520)* | | (-4.297) |
| *Choice probability elasticities* | | | | | | |
| Travel time (minutes) | 0.000 | -13.635 | -5.145 | -0.561 | -5.203 | -5.075 |
| Travel cost ($) | 0.000 | 0.000 | -0.116 | -0.035 | 0.000 | -0.306 |
| *Marginal rates of substitution* | | | | | | |
| Value of time ($/hr) | - | ∞ | 19.315 | 34.101 | ∞ | 7.376 |

*Not Significant at 95% level

**T-test estimates are presented in the parenthesis.

**Individual Class 6:** This class constitutes 2% of the sample population. Individuals belonging to this class consider all travel modes except private transit and bike for mandatory tours, and all travel modes except private transit for non-mandatory tours. Like Class 5, travel mode shares for private vehicle are low, with 23% of mandatory tours and 57% of non-mandatory tours being made by that mode. Unlike Class 5, the class has a low value of time of 6.8$/hr and 7.4$/hr for mandatory and non-mandatory tours, respectively. Walk is by far the most popular alternative mode of transport, accounting for 40% of mandatory tours and 35% of non-mandatory tours. Public transit usage for mandatory tours is quite high as well, at 38%, but the corresponding figure for non-mandatory tours is much lower at 7%. The class comprises a mix of adult parents and children still in school.

## 5.2 Household modality style estimates:

As mentioned previously, the final specification for the household residential location choice behaviour, identified six classes in the sample population that differ from each other in terms of the household demographic characteristics and neighbourhood attributes. In table 5 we presents the estimated results for the household class membership model having segment 1 as the base for our estimate. Table 6 provides estimation results per class for the neighbourhood choice model. Below we summarize some of the key attributes of each of these six classes.

**Household Class 1:** This class constitutes 10.3% of the sample population, and consists largely of young urban professionals with multimodal lifestyles.

Households belonging to this class have low levels of both home ownership (38%) and car ownership (0.88 cars per household) and they are less likely to reside in a single family house (12%). They have small households (average household size of 1.90) that are unlikely to include children (16% of these households have children), and high household incomes ($122,000 per year), indicating high per capita incomes.

In terms of neighbourhood location, households belonging to this class prefer neighbourhoods that are dense, diverse and walkable (as indicated by a positive coefficient on the design variable, i.e. blocks per square mile), such as central business districts (CBDs). The class is more likely to choose neighbourhoods with higher rents and property values. While this might seem counter intuitive, we wish to remind the reader that rents and property values are being used as proxies for neighbourhood quality, as denoted by variables not explicitly included in the class-specific model specification, such as crime, cleanliness and local amenities (Ardeshiri, Ardeshiri et al. 2016).

As one would expect, individuals belonging to this household class are much more likely themselves to belong to one of the two individual classes with low car dependence (individual classes 5 and 6).

**Household Class 2:** This class constitutes 28% of the sample population and consists largely of affluent professionals with partial car oriented lifestyles.

Households belonging to this class have the highest level of both home ownership (96%) and bicycle ownership (2.55 bicycles per household) and second highest car ownership (2.12 cars per household). On average they have a household size of 2.78 and have the highest level of household incomes ($160,000 per year).

In terms of neighbourhood location households belonging to this class prefer neighbourhoods that have low density and are walkable. On average the occupied population are whites (69%). The class is more likely to choose neighbourhoods with highest rents and property values and as previously mentioned this means this class are seeking for high neighbourhood quality.

Individuals belonging to this household class are much more likely themselves to belong to one of the two classes with partial car dependence (individual classes 2 and 3).

**Household Class 3:** This class constitutes 31.2% of the sample population and consists largely of traditional households with complete car oriented lifestyles.

Households belonging to this class have the highest level of car ownership (2.19 cars per household) and almost all (99%) live in an owned (95%) single family household. On average they have a household size of 2.55 with total annual income of $93,000. These households are highly likely to be white Americans.

With regard to the neighbourhood location households belonging to this class prefer neighbourhoods that have lowest density and diversity measure among all other neighbourhood. Its race-ethnic composition are mainly whites (68%) Hispanics from the middle class group. Median Rent and property value are relevantly lower in this class in comparison with classes 1 and 2.

Individuals belonging to this household class are much more likely themselves to belong to complete car dependence class (individual class 1).

**Household Class 4:** This class constitutes 8.1% of the sample population and consists largely of Hispanics with partial car oriented lifestyles.

Households belonging to this class have on average 1.65 cars. Households in this class have the highest household size with 3.56 with the lowest total annual income of $44,000. They are more likely to be Hispanic (79% on average).

In terms of neighbourhood location, none of the 3 D's measures have become significant for this class. In addition the median property and rent value has relatively the lowest amount. This Indicates that neighbourhood's affordances has a prior to neighbourhood's quality when it comes to choosing a neighbourhood location to reside.  Low class income with Hispanic background are the dominant population in neighbourhoods.

Although individuals belonging to this household class coming from a low class income however, they are more likely themselves to belong to one of the two classes with partial car dependence (individual classes 2 and 3).

**Household Class 5:** This class constitutes 15% of the sample population and consists largely of middle aged actives with multimodal lifestyles.

Subsequently to households from class 1, this class have the second lowest car ownership (1.19 cars per household) and are less likely to reside in a single family house (14%). On the contrary, this households on average own the second highest number of bikes (1.73 per household).  On average they have a household size of 1.74 and the total household income is $59,000 per year.

In terms of neighbourhood location households belonging to this class prefer neighbourhoods that are diverse and walkable, which provides the opportunity to be less car oriented and more open to other travel modes such as cycling or walking. This argument can also be supported by the fact that households in this class own high number of bicycles.

Individuals belonging to this household class are much more likely themselves to belong to one of the two individual classes with low car dependence (individual classes 5 and 6) or the multimodal individual class (individual class 4).

**Table 5.** Estimated results for the household class membership model

| Parameters | Segment 2 | Segment 3 | Segment 4 | Segment 5 | Segment 6 |
|---|---|---|---|---|---|
| Constant | -5.548<br>(-5.152) | -3.695<br>(-3.265) | 2.666<br>(2.365) | 3.919<br>(5.430) | 4.274<br>(4.838) |
| Number of vehicles | 4.574<br>(32.732) | 5.244<br>(37.765) | 4.843<br>(16.3) | 3.017<br>(12.846) | 4.645<br>(19.69) |
| Driver's license holding | -0.505<br>(-2.138) | -0.602<br>(-2.643) | -2.238<br>(-5.551) | -0.794<br>(-2.544) | -0.623*<br>(-1.705) |
| Number of bikes | 0.083<br>(2.522) | -0.232<br>(-3.785) | -0.653<br>(-3.862) | 0.1363<br>(3.709) | -0.49<br>(-4.064) |
| Single Family House | 7.198<br>(16.15) | 9.100<br>(12.04) | 4.857<br>(10.111) | 1.759<br>(5.499) | 3.097<br>(7.854) |
| Residence Owned | -1.549<br>(-3.438) | -2.892<br>(-6.964) | -1.768<br>(-3.416) | -0.715<br>(-3.703) | -2.672<br>(-6.064) |
| Annual household income | -0.269*<br>(-1.126) | -3.721<br>(-14.88) | -8.375<br>(-11.35) | -4.842<br>(-12.36) | -1.991<br>(-6.985) |
| Household size | 0.4759<br>(2.102) | 0.5071<br>(2.471) | 3.0547<br>(9.155) | 0.3288*<br>(1.153) | 1.3625<br>(4.684) |
| Residence tenure (years) | -0.027<br>(-2.699) | -0.136<br>(-14.27) | -0.148<br>(-8.55) | -0.129<br>(-9.72) | -0.116<br>(-6.959) |
| Number of workers | -1.866<br>(-13.59) | -1.419<br>(-10.9) | -1.758<br>(-6.52) | -1.107<br>(-5.214) | -1.494<br>(-6.508) |
| Number of students | 0.612<br>(2.831) | 0.306*<br>(1.576) | -0.977<br>(-3.082) | 0.0235*<br>(0.084) | -0.584<br>(-1.991) |
| Unrelated household | -0.698*<br>(-1.288) | -1.702<br>(-3.246) | -0.937*<br>(-1.457) | 0.208*<br>(0.367) | -2.565<br>(-1.195) |
| Pre-school kids | -2.175<br>(-4.256) | -1.913<br>(-4.116) | -2.447<br>(-3.326) | -0.808*<br>(-1.305) | -0.689*<br>(-1.26) |
| School kids | -0.613<br>(-1.669) | -1.05<br>(-3.004) | -1.013<br>(-1.46) | -1.204<br>(-2.026) | 1.296<br>(2.336) |
| Hispanic household | -2.003<br>(-5.351) | -0.13*<br>(-0.461) | 5.739<br>(7.657) | 0.358*<br>(0.938) | -0.73*<br>(-1.589) |
| White house hold | 1.232<br>(3.648) | 2.273<br>(6.409) | -3.683<br>(-6.945) | -0.134*<br>(-0.416) | -4.014<br>(-9.333) |
| Black household | -0.241*<br>(-0.171) | 4.262<br>(7.573) | 8.351<br>(12.14) | 4.848<br>(9.722) | 3.204<br>(5.520) |
| American household | -0.534*<br>(-1.013) | 1.404<br>(2.395) | -1.484<br>(-2.798) | 1.734<br>(3.516) | -2.299<br>(-5.184) |

*Not Significant at 95% level
**T-test estimates are presented in the parenthesis.

**Household Class 6:** This class constitutes 7.3% of the sample population and consists largely of Immigrants with complete car oriented lifestyles.

Households belonging to this class have on average 2.05 cars and 86% of them are home ownerships. With a household size of 3.05 they are the second largest family size with a total household incomes of $125,000 per year.

In terms of neighbourhood location, similar to class 4 none of the 3 D's variable are significant for this class. Immigrants constitute the largest population group in these neighbourhoods.

Similar to class 3, individuals belonging to this household class are much more likely themselves to belong to complete car dependence class (individual class 1).

**Table 6.** Class specific for neighbourhood choice model

| Parameters | Class 1 | Class 2 | Class 3 | Class 4 | Class 5 | Class 6 |
|---|---|---|---|---|---|---|
| **Density** | 0.457 | -4.161 | -5.715 | -0.578* | -0.252* | -0.585* |
| (100,000 per square mile) | (4.017) | (-8.829) | (-7.424) | (-1.637) | (-1.498) | (-0.888) |
| **Diversity** | 0.521 | -0.027* | -0.239 | 0.046* | 0.506 | 0.041* |
| (Entropy) | (6.224) | (-0.477) | (-4.650) | (0.573) | (8.491) | (0.427) |
| **Design** | 0.765 | 0.313 | 0.117* | -0.096* | -0.001* | -0.181* |
| (Number of census blocks per square mile) | (13.05) | (5.246) | (1.621) | (-1.343) | (-0.020) | (-1.485) |
| **Whites** | 3.105 | 1.039 | 0.809 | -0.907 | 2.629 | -1.700 |
| | (8.203) | (4.541) | (4.578) | (-3.234) | (12.11) | (-3.760) |
| **Blacks** | 2.222 | -2.894 | -1.152 | 0.848 | 1.472 | 0.359* |
| | (3.535) | (-3.908) | (-3.600) | (2.327) | (4.269) | (0.397) |
| **Hispanics** | -1.192 | 0.056* | 0.612 | 3.477 | -2.667 | -1.989 |
| | (-3.154) | (0.198) | (2.756) | (11.42) | (-10.26) | (-3.676) |
| **Immigrants** | 3.035 | -0.114* | -2.742 | -0.386* | 2.184 | 2.702 |
| | (6.761) | (-0.321) | (-8.118) | (-0.822) | (6.297) | (4.198) |
| **Upper Class** | -2.809 | 2.102 | -0.084* | -0.931* | -4.409 | 2.277 |
| | (-4.384) | (5.976) | (-0.234) | (-1.138) | (-7.655) | (3.107) |
| **Upper Middle Class** | 2.366 | 2.189 | 2.284 | 0.426* | -2.594 | 3.617 |
| | (2.730) | (4.613) | (5.632) | (0.494) | (-4.096) | (4.293) |
| **Lower Middle Class** | 0.871* | 1.991 | 0.669* | 0.073* | 1.028 | 3.713 |
| | (1.216) | (4.176) | (1.774) | (0.115) | (2.143) | (3.975) |
| **Median Rent** | 0.651 | 0.829 | -0.220 | -0.136* | -0.269* | 0.350* |
| | (3.024) | (-2.457) | (-2.341) | (-0.668) | (-1.722) | (1.675) |
| **Median Value** | 2.833 | 2.883 | -0.371 | -0.877 | 0.494 | 0.335* |
| | (11.02) | (10.81) | (-1.964) | (-2.588) | (2.310) | (0.964) |

*Not Significant at 95% level
**T-test estimates are presented in the parenthesis.

In order to provide readers with a better description of the outcomes mentioned above, we prepared maps images for two counties from the bay area to present the household modality style when it comes to selecting a neighbourhood to live base on belonging to one of the six household segments. Figure 2 and 3 respectively presents the expected neighbourhood choices for San Francisco and San Jose with respect to the household segments. Dark blue area represents the most desired census tracts where the white tract zones represent the least preferred tract zones. For example, households belonging to class 1 prefer to reside at north east of San Francisco where the CBD and the financial district is located or household belonging to class 2 prefer to live at tracts located at the heart of San Francisco (e.g. Ingleside Terraces and Westwood Highlands area).

**Figure 2.** Expected neighbourhood choices for San Francisco Area with respect to the household segments.

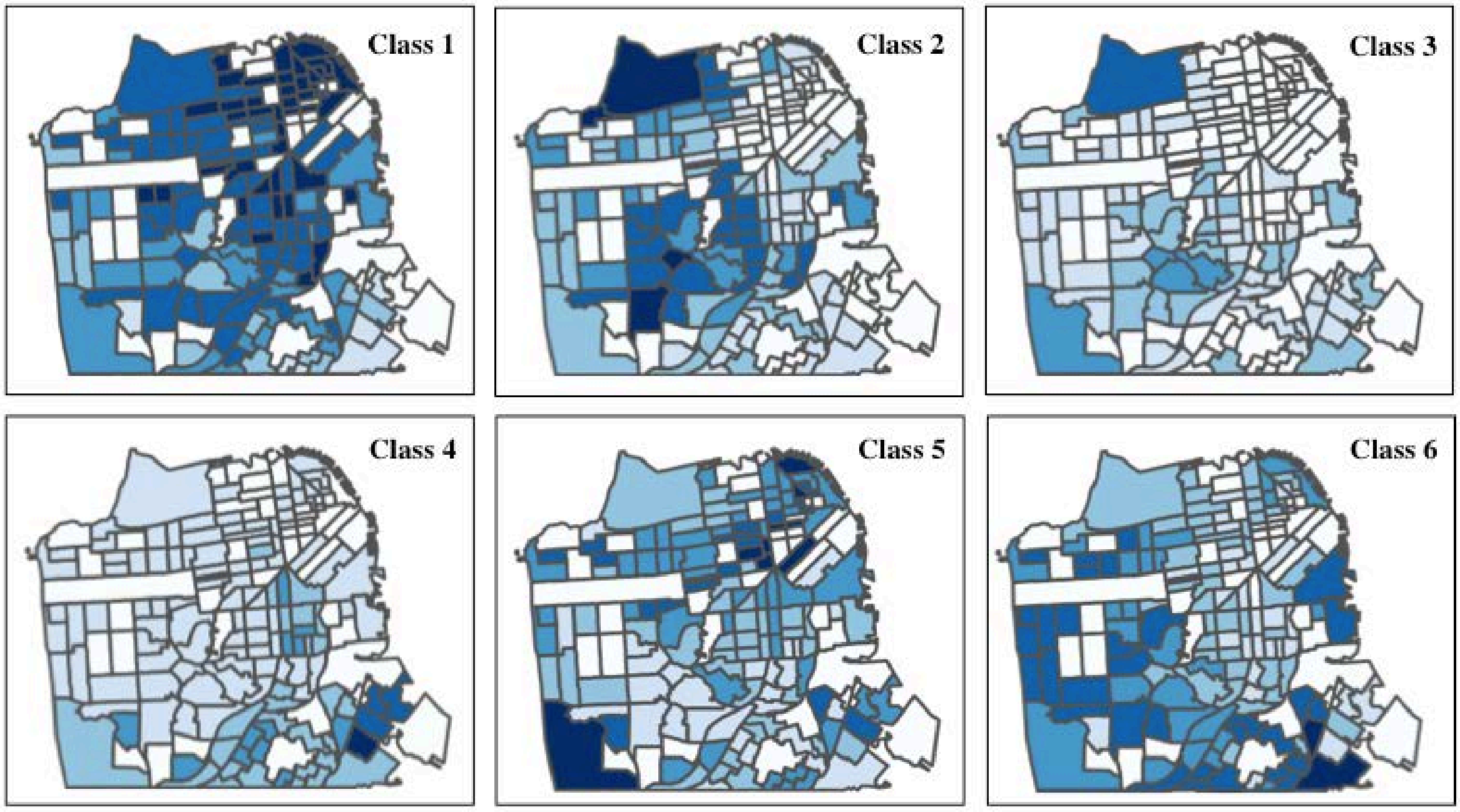

**Figure 3.** Expected neighbourhood choices for San Jose Area with respect to the household segments.

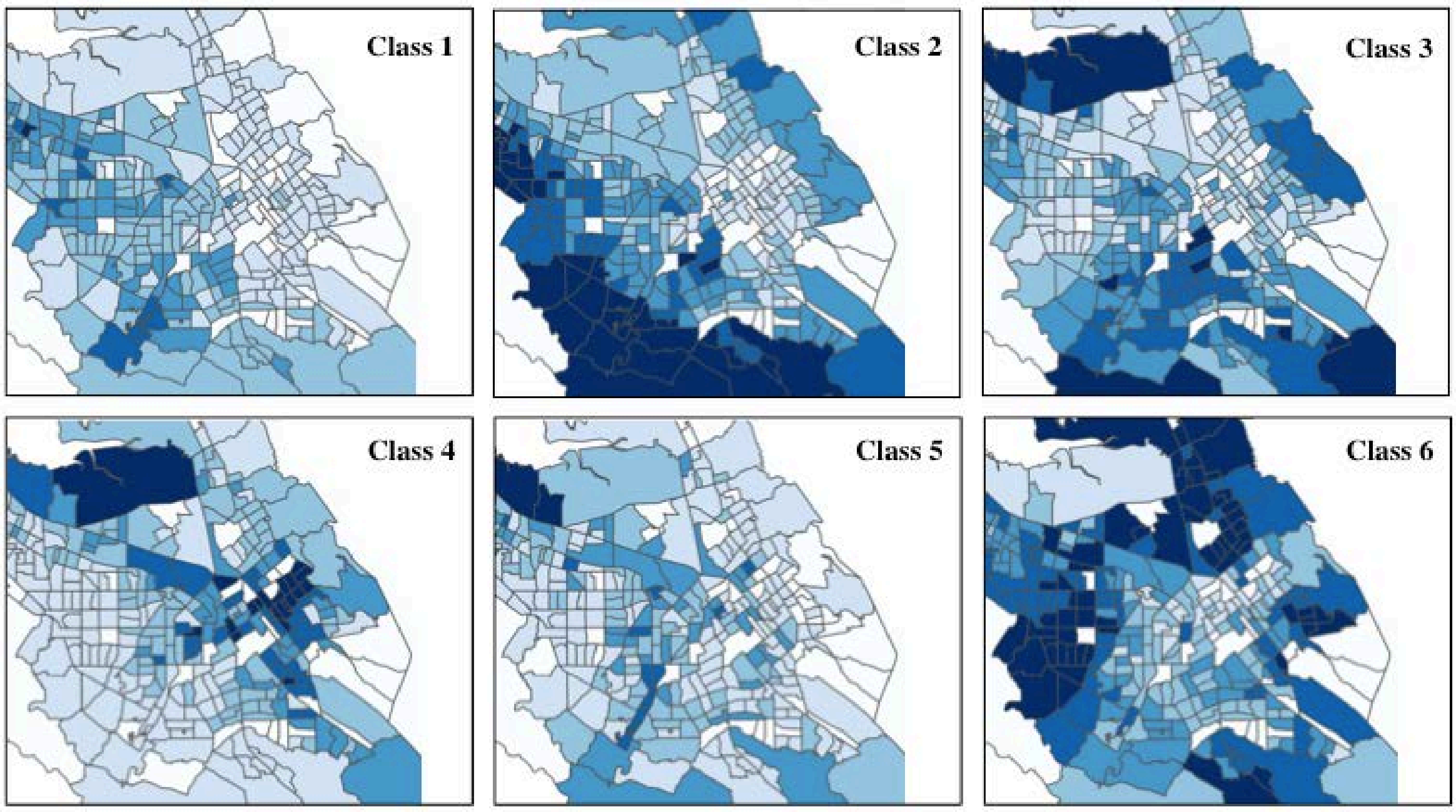

**Table 7.** Estimation results for the class membership model corresponding to individual modality styles, conditional on the household modality style

| Parameters | Class 1 | Class 2 | Class 3 | Class 4 | Class 5 | Class 6 |
|---|---|---|---|---|---|---|
| Class-specific constant (Class 2) | -0.339 | 1.838 | 3.107 | -31.993 | -0.034 | 3.306 |
| Class-specific constant (Class 3) | 0.661 | 3.161 | 2.767 | 2.156 | 1.835 | 0.402 |
| Class-specific constant (Class 4) | -1.449 | -59.552 | -69.751 | -59.554 | -0.100 | -68.138 |
| Class-specific constant (Class 5) | 0.429 | 1.374 | 1.076 | -1.240 | -0.101 | -0.139 |
| Class-specific constant (Class 6) | 1.003 | 2.986 | 3.105 | 3.859 | 2.248 | 3.201 |
| Male (Class 2) | -0.469 | 0.334 | 0.565 | 13.097 | -36.377 | -0.285 |
| Male (Class 3) | -0.663 | -0.023 | -0.209 | 3.006 | -0.296 | 2.420 |
| Male (Class 4) | 1.806 | 15.529 | 5.921 | 15.528 | 0.713 | 6.066 |
| Male (Class 5) | -0.844 | 0.110 | 0.746 | -30.850 | -0.006 | -27.480 |
| Male (Class 6) | 0.147 | -0.048 | 0.296 | 0.615 | -0.004 | -0.594 |
| Married (Class 2) | -1.029 | 2.119 | 0.925 | -59.549 | -33.676 | 1.592 |
| Married (Class 3) | -0.555 | -0.179 | -30.853 | -1.068 | 0.574 | -0.688 |
| Married (Class 4) | 0.339 | 15.534 | 5.129 | 15.533 | -0.595 | 2.997 |
| Married (Class 5) | -0.466 | -0.208 | -0.546 | -35.796 | 0.269 | -0.912 |
| Married (Class 6) | -0.002 | 0.133 | 0.275 | -0.582 | 0.308 | 0.989 |
| Parent (Class 2) | -30.848 | 1.557 | 0.924 | 13.570 | -34.208 | 2.157 |
| Parent (Class 3) | -1.593 | 1.043 | 1.031 | 1.668 | 1.259 | -30.846 |
| Parent (Class 4) | -30.847 | 15.533 | 11.606 | 15.530 | 0.457 | 7.698 |
| Parent (Class 5) | -30.854 | -59.549 | -0.096 | 3.462 | 0.434 | -0.421 |
| Parent (Class 6) | -0.417 | 0.278 | 0.414 | 1.933 | 0.763 | 0.526 |
| Employed (Class 2) | 0.475 | 1.198 | 0.043 | -59.550 | -0.796 | 1.500 |
| Employed (Class 3) | 0.355 | 0.454 | 1.672 | -0.180 | 0.274 | 1.229 |
| Employed (Class 4) | -0.175 | 15.532 | 7.907 | 15.535 | 0.680 | -30.024 |
| Employed (Class 5) | 0.171 | -0.109 | -0.148 | -0.240 | 1.070 | -31.421 |
| Employed (Class 6) | 0.164 | 0.076 | 0.582 | -1.815 | 0.057 | 0.678 |
| Student (Class 2) | -59.550 | -0.529 | 1.368 | -12.486 | -30.847 | 55.892 |
| Student (Class 3) | 1.533 | -1.356 | -30.845 | -4.037 | -0.323 | 15.537 |
| Student (Class 4) | 1.150 | 15.531 | -32.672 | 15.531 | 0.453 | 15.533 |
| Student (Class 5) | -59.546 | -59.550 | 0.179 | -30.853 | 1.615 | 15.532 |
| Student (Class 6) | 2.007 | 0.415 | 0.242 | -1.113 | 1.023 | 15.535 |
| Age (Class 2) | -0.011 | -0.032 | 0.002 | -30.845 | 0.022 | -0.047 |
| Age (Class 3) | 0.022 | 0.007 | -0.021 | 0.005 | 0.010 | 0.035 |
| Age (Class 4) | 0.010 | -2.762 | 0.648 | -2.126 | 0.010 | 0.712 |
| Age (Class 5) | 0.015 | 0.010 | 0.015 | 0.015 | 0.001 | 0.067 |
| Age (Class 6) | -0.001 | 0.009 | 0.007 | -0.006 | 0.001 | -0.005 |

## 6. Conclusion and further discussions

This study proposed a methodological framework to capture the concurrent influence of lifestyles and modality styles on household neighbourhood location and individual travel mode choice behaviours. The framework was empirically tested using travel diary data collected from households residing in the San Francisco Bay Area, United States.

There are a number of benefits to the model framework over more traditional representations of individual and household travel and activity behaviour. First, the horizontal framework stands in stark contrast to vertical representations usually employed by activity-based travel demand models currently in practice, such as the San Francisco Chained Activity Modelling Process (SF-CHAMP; Cambridge Systematics, 2002), where lower order dimensions, such as travel mode choices, are conditioned on higher order dimensions, such as neighbourhood location. The horizontal representation serves to emphasize the absence of hierarchies between different dimensions of transportation and land use behaviour and the presence instead of correlation across all dimensions, induced through the modality styles construct.

Second, the framework is more nuanced in its representation of the relationship between the built environment and travel behaviour. In particular, it controls for residential self-selection by allowing households with different modality styles to self-select into neighbourhoods that best serve their needs. For example, the framework identified six household classes in the Bay Area that differ in their predisposition towards different land use measures, such as density and diversity, and their proclivity to use different travel modes.

And finally, the model captures the dynamic underlying group decision-making through the interplay between household and individual modality styles. By conditioning individual modality styles on household modality styles, we allow individual preferences to be some reflection of the preferences of the household as a whole. At the same time, individual circumstances, such as age and employment, can potentially override the influence of household modality styles.

## Appendix A

Variables used to model our residential neighbourhood choice and individual travel mode choices

| Neighbourhood variables | Description |
|---|---|
| Density | Number of residents and employees per square mile. |
| Diversity | An entropy measure calculated using the proportions of the population that are residents or employees belonging to one of fourteen different industries |
| Design | Number of census blocks per square mile |
| Whites | Proportion of whites residents in the census tract. |
| Blacks | Proportion of blacks residents in the census tract. |
| Hispanics | Proportion of Hispanics residents in the census tract. |
| Immigrants | Proportion of Immigrants residents in the census tract. |
| Upper Class | Proportion of upper class residents in the census tract. |
| Upper Middle Class | Proportion of upper middle residents in the census tract. |
| Lower Middle Class | Proportion of lower middle class residents in the census tract. |
| Median Rent | Median rent of a residential unit in the census tract. |
| Median Value | Median value of a residential unit in the census tract. |
| **Household variables** | **Description** |
| Number of vehicles | Number of vehicle/s owned in the household. |
| Number of bikes | Number of bike/s owned in the household. |
| Single Family House | A 0-1 dummy variable representing if the household residential type is a single family house. |
| Home Owner | A 0-1 dummy variable indicating if the dwelling is owed by the household. |
| Annual household income | The total annual household income. |
| Household size | A measure representing the household size. |
| Residence tenure | A 0-1 dummy variable indicating if the dwelling under a residence tenure. |
| Number of workers | An indicator of number of individual contributing to the total household income. |
| Number of students | An indicator of number of students in the household. |
| Driver's license holding | An indicator of number of households that hold a driver's license. |
| Unrelated household | A 0-1 dummy variable indicating if household members are related to each other. |
| Pre-school kids | An indicator representing the number of pre-school kids in the household. |
| School kids | An indicator representing the number of school kids in the household. |
| Hispanic household | A 0-1 dummy variable indicating if the household is Hispanic. |
| White house hold | A 0-1 dummy variable indicating if the household is white. |
| Black household | A 0-1 dummy variable indicating if the household is black. |
| American household | A 0-1 dummy variable indicating if the household is native American. |
| **Travel mode variables** | **Description** |
| Travel mode | The travel modes examined for this study are car, private transit, public transit, bike and walk. |
| Travel time | The travel time base on mode specific. |
| Travel cost | The travel cost base on mode specific. |
| **Individual variables** | **Description** |
| Male | A 0-1 dummy variable indicating if the individual is male. |
| Married | A 0-1 dummy variable indicating if the individual is married. |
| Parent | A 0-1 dummy variable indicating if the individual is a parent. |
| Employed | A 0-1 dummy variable indicating if the individual is employed. |
| Student | A 0-1 dummy variable indicating if the individual is student. |
| Age | An indicator representing the individual's age. |